\documentclass{aa}

\usepackage{graphicx}

\newcommand{\kmsec}{\mbox{$\rm{km \; s^{-1}}$}}

\newcommand{\jnu}{\mbox{$j_\nu$}}
\newcommand{\mH}{\mbox{$m_{\rm H}$}}

\newcommand{\msun}{\mbox{$M_\odot$}}
\newcommand{\mdot}{\mbox{$\dot{M}$}}

\newcommand{\vinf}{\mbox{$v_\infty$}}

\newcommand{\be}{\begin{equation}}
\newcommand{\ee}{\end{equation}}

\newcommand{\kappaw}{\mbox{$\kappa_{\rm w}$}}
\newcommand{\Lx}{\mbox{$L_{\rm X}$}}
\newcommand{\Tx}{\mbox{$T_{\rm X}$}}
\newcommand{\fx}{\mbox{$f_{\rm X}$}}
\newcommand{\tm}{\mbox{${\rm t}$}}
\newcommand{\nel}{\mbox{$n_{\rm e}$}}
\newcommand{\nion}{\mbox{$n_{\rm i}$}}

\def \etal\,{et~al.\/}
\def\lesssim{\mathrel{\hbox{\rlap{\hbox{\lower4pt\hbox{$\sim$}}}\hbox{$<$}}}}
\def\gtrsim{\mathrel{\hbox{\rlap{\hbox{\lower4pt\hbox{$\sim$}}}\hbox{$>$}}}}

\def\figurenum#1{\def\thefigure{#1}\let\@currentlabel\thefigure
\addtocounter{figure}{\m@ne}}
\def\figcaption{\@ifnextchar[{\@xfigcaption}{\@figcaption}}
\def\@figcaption#1{{\def\@captype{figure}\caption{#1}}}

\def\@xfigcaption[#1]#2{{\def\@captype{figure}\caption{#2}}}
\def\fnum@figure{{\rm Figure\space\thefigure:}}
\def\fps@figure{bp}
\def\eps@scaling{.95}\def\epsscale#1{\gdef\eps@scaling{#1}}
\def\plotone#1{\centering \leavevmode
\epsfxsize=\eps@scaling\columnwidth \epsfbox{#1}}

\begin{document}
\thesaurus{07           
           (
           08.01.1;     
           08.05.1;     
           08.13.2;     
           08.23.2;     
           13.25.5      
           )}

\title{Modelling X-ray variability in the structured atmospheres of hot
stars }
\author{L.~M.~Oskinova$^1$,\, R.~Ignace$^2$,\, J.~C.~Brown$^1$,\, J.~P.
Cassinelli$^3$}
\institute{$^1$ Department of Physics and Astronomy, University of
Glasgow, Glasgow, G12 8QQ, Scotland UK
\\~
$^2$ Department of Physics and Astronomy, University of Iowa, 203 Van Allen Hall,
 Iowa City, IA 52242, USA \\~
$^3$ Department of Astronomy, University of
 Wisconsin, 5534 Sterling
Hall, 475 N. Charter St., Madison, WI  53706-1582, USA
}

\offprints{lida@astro.gla.ac.uk}
\date{Received <date>; Accepted <date>}
\authorrunning{Oskinova \etal\,}
\titlerunning{X-ray variability modelling}

\maketitle
\begin{abstract}

We describe X-ray production in the atmospheres of hot, early-type
stars in the framework of a ``stochastic shock model''.  The extended
envelope of a star is assumed to possess numerous X-ray emitting
``hot'' zones that are produced by shocks and embedded in the ambient
``cold'' medium in dynamical equilibrium. It is shown that the apparent
lack of X-ray variability on short ($\sim$ hours) timescales do not
contradict a shock model for X-ray production. The character of the
X-ray variability is found to depend on the frequency with which hot
zones are generated, the cool wind opacity to X-rays, and the wind flow
parameters, such as mass loss rate and terminal speed.

\end{abstract}
\section{Introduction}

The X-ray emission from hot stars has proved to be an important
``window'' for investigating mass loss in early type stars, and the
advent of the latest suite of orbital X-ray telescopes are providing
new insights.  Here we develop a model to explore the X-ray variability
of early type stars, first to understand the near absence of
significant variability observed so far, and second to determine the
requisite S/N and time resolutions necessary to measure the expected
X-ray variability with current and future instrumentation.  The objects
of interest are hot, luminous stars with supersonic winds that carry
significant mass, such as OB and Wolf-Rayet (WR) stars.  The main
parameters characterising these stellar winds are the terminal velocity
(\vinf) and mass loss rate (\mdot). Typical O star mass loss rates are
$\mdot \approx 10^{-7}$\msun yr$^{-1}$ with terminal speeds $\vinf
\approx 1000-3000$\,\kmsec.  For Wolf-Rayet (WR) stars \mdot\ is larger
at around $10^{-4.5}$\msun yr$^{-1}$ but with terminal speeds like
those of O~stars. Such strong outflow results in a predominance of
anomalously strong and broad emission lines in the spectra of
WR~stars.  The high mass loss rate during the WR~phase has a
significant effect on its evolution and the chemical enrichment and
mechanical energy input to the interstellar medium.

It is widely accepted now that stellar winds of OB~stars are maintained
by radiation pressure in numerous lines of heavy elements (Castor \etal\,
1975; Pauldrach \etal\, 1986). But the acceleration of
high mass loss WR~winds is though to be due to multi-line scattering of
photons (Lucy \& Abbott 1993; Springmann 1994; Gayley \etal\, 1995).
Non-LTE, moving atmosphere modelling codes have been developed (e.g.,
Hamann \& Koesterke 2000; Hillier \& Miller 1998), which adopt
radiative and statistical equilibrium  and assume a monotonic velocity
law {\it a priori}. In the framework of the so-called ``standard
model'' the emission lines are formed in a spherically symmetric,
time-independent, dense, smooth stellar wind that is photoionized by
the hot core.  However, there has long been growing evidence that at
least some assumptions of the standard model are oversimplified.  A
pointed example is the X-ray emission from hot star winds, first
detected as discrete X-ray sources with {\sc Einstein} (0.2--4.0 keV)
(Harnden \etal\, 1979; Seward \etal\, 1979; Seward \& Chlebowski 1982).

Early UV observations of hot stars revealed the appearance of
superionisation, especially the presence of O{\sc vi} which would not
be expected in the winds of hot stars.  Cassinelli \& Olson (1979)
suggested that Auger ionisation by X-rays in the context of a coronal
model could explain the existence of these highly ionised species.
However, the hypothesis of a hot corona in the outer parts of a stellar
atmosphere was ruled out due to the fact that the soft X-ray flux is
not sufficiently absorbed at low energies (e.g., Cassinelli \& Swank
1983). According to a phenomenological model proposed by Lucy \& White
(1980) and developed by Lucy (1982), dynamical instabilities should
arise in line driven stellar winds leading to shock generation
throughout the flow.  Cassinelli \& Swank (1983) found that the shock
model of Lucy (1982) produced too low an X-ray flux and also that the
X-rays were too soft to explain the {\sc Einstein} observations of the
O~stars.  They also pointed out that the spherically symmetric shock
model would lead to strong variability of the X-rays, which was not
seen.  To explain the lack of clear evidence for variability of the
X-rays from hot stars they proposed that the X-rays are not from
spherically symmetric shocks but instead from shock fragments in
the wind.  
{\sc Rosat} (0.2--2.4 keV) observations have brought new insight in the
properties of X-ray emission from hot stars.  Kudritzki \etal\, (1996)
have found interesting relations between the X-ray emission and the
wind properties from an analysis of 42 O~star spectra.   For example,
they confirm the relation between X-ray luminosity and stellar
bolometric luminosity as $L_{\rm X} \approx 10^{-7}\,L_{\rm Bol}$ that
was previously known, and also find that $L_{\rm X}$ scales with the
wind kinetic luminosity $0.5 \mdot v_\infty^2$.  For the WR stars, the
data is of lower quality and consists almost entirely of passband
fluxes.  However, based on {\sc Einstein} observations, Pollock (1987)
was able to find that on average, N-rich WR stars (WN) tend to be more
luminous than the C-rich WR stars (WC), possibly attributable to the
difference in abundances of the two types.  The {\sc Rosat} data have
also revealed that, unlike their progenitors the O~stars, the X-ray
luminosities of single WR stars are {\it not} correlated with $L_{\rm
Bol}$, wind momentum \mdot \vinf, wind kinetic luminosity $0.5 \mdot
v_\infty^2$, or WR subtype (Wessolowski 1996; Ignace \& Oskinova
1999).  Recent data from {\sc Chandra} and {\sc XMM-Newton} of OB stars
(Schulz \etal\, 2000; Kahn \etal\, 2001; Waldron \& Cassinelli 2001)
and WR stars (Maeda \& Tsuboi 1999) are just becoming available.
These newer data, mostly of O~stars, are showing rich X-ray
line spectra, the analysis of which are yielding some unexpected
results.  We refer the reader to the above mentioned papers for
a description of these data and here concentrate on the topic
of X-ray variability in single stars, a topic that has not
much been addressed from a theoretical perspective.

Dynamical instabilities in line-driven winds have been studied
extensively for OB stars (Owocki \& Rybicki 1984; Owocki \etal\, 1988;
Rybicki \etal\, 1990).  Model computations predict shock velocity-jumps
ranging from 500 km s$^{-1}$ to 700 km s$^{-1}$, implying post-shock
temperatures and emission measures that are marginally adequate to
explain the observed thermal X-ray emission (e.g. Hillier \etal\, 1993,
Feldmeier \etal\, 1997a).  Gayley \& Owocki (1995) have shown that
similar to OB~stars, the instability mechanism should also operate in
the denser WR~ winds.

In terms of observed X-ray variability, Crowther \& Willis (1996) have
presented a review of available data based on {\sc Rosat} observations.
For $\zeta$~Ori, Bergh{\"o}fer \& Schmitt (1994) claimed to detect
X-ray variability over a 2-day period, reaching about 30\% in hard
X-rays (0.64-2.38 keV), while the soft band data (0.16-0.5includegraphics1 keV)
remained  about constant. For $\zeta$ Pup, Bergh{\" o}fer \etal\,
(1996) found evidence for a modulation in the harder (0.9--2.0 keV)
X-ray emission and in the optical data with a 16.7 hour period, and
suggested that this reflects periodic changes in the base wind density
of the O4~f star. On the other hand, the study of $\sigma$ Ori showed
no significant X-ray variability.  Except for WR\,6 (HD\,50896) and
WR\,1 (HD\,4004), no X-ray variability studies for single WR~stars have
been undertaken. For WR~6, Willis \& Stevens (1996) reported
variability at the $\le$ 30\% level on timescale of $\le 1$~day,
together with larger epoch changes.  No significant spectral shape
changes were found.  In the case of WR~1, no evidence of significant
X-ray variability was reported by Wessolowski \& Niedzielski (1996).
Thus far, no short-time scale X-ray variability on the order of the
wind flow time ($R_*/\vinf\sim$ hours) has been observed in O and
WR~stars.

In this paper we consider the application of stochastic inhomogeneous
wind models, as widely accepted necessary components to explain
spectral variations at optical and UV wavelengths, to the X-ray
emission from single O and WR~type stars. We restrict our analysis to
shock models of X-ray formation in hot stars and do not consider
alternative models (see discussion in Waldron \& Cassinelli 2001). In
Sect.\,2, we describe a stellar wind model consisting of numerous zones
filled by hot X-ray emitting gas. Two following sections use two
different approximations of the optical depths of the cool material,
which absorbs X-ray emission.  In Sect.\,3 we apply the so-called
exospheric approximation treating the X-ray emitting zones as spherical
shells.  In Sect.\,4 we generalise our model by using formal
integration of the radiative transfer equation for randomly distributed
wind shocks.  We investigate the crucial parameters of the variability
and their relation to the input parameters of the model.  A summary
discussion of the results is presented in Sect.\,5.

\section{Stochastic model for the distribution of X-ray emitting material.}

We consider a spherically symmetric and time\,-independent stellar wind
that is a mix of ``cool'' and ``hot'' gas in dynamical equilibrium. The
minor hot gas component gives rise to X-ray emission. We assume that
hot gas is present in the form of spatially separated compact volumes
characterised by temperature \Tx\, and density which are different from
the temperature and density of the ambient cool wind gas.

It is common in the literature to attribute wind ``clumps'' to dense
formations with temperatures much lower then the temperatures of the
X-ray emitting material, sometimes referred to as DWEEs (discrete wind
emission elements, L{\'e}pine \& Moffat 1999).  These clumps are
presumably responsible for observed line profile variability in the
optical spectra of WR~stars, or the formation of DACs (discrete
absorption components) observed in the optical and UV spectra of some
O~stars (Brown \etal\, 1995).  Such cold clumps and hot X-ray emitting
zones may well have related origins.  Feldmeier \etal\, (1997a)
considered the formation of X-ray emitting zones as a consequence of
shell collisions due to the propagation of reverse shocks in the
stellar wind. In their hydrodynamical simulations they demonstrated the
existence of dense shells moving according to the stationary wind
velocity law and small-mass, high-velocity blobs. Copious X-rays are
produced when a fast blob rams into an outer shell.  These X-rays thus
originate over a fairly narrow range in radius.  In this respect,
recent observations (Schulz \etal\, 2000) seem to confirm such a
scenario, except that the expected strong variability is not observed.

The general description  of a source with a continuous spatial
distribution of temperature (e.g., a radiative cooling zone behind the
front of a shock wave) is given by the {\it differential emission
measure} which includes integration  over all  surfaces with the same
temperature, such as over multiple independent shocks in a stellar wind
(Craig \& Brown 1976). But it is necessary to note that an overall
consistent picture of dynamical, structured, thick stellar winds is
still far from being complete. Although detailed  models do not predict
the X-ray sources to be isothermal with unique temperature  \Tx\,
amongst them, the isothermal approximation for the overall wind
structure should be fairly close to the actual case because of the
short radiative cooling times in these dense stellar winds (Feldmeier
\etal\, 1997b).  The two-temperature approximation  with isothermal
X-ray emitting zones embedded in a cooler outflowing wind is able to
reasonably well reproduce the observed X-ray spectra from hot stars
(Hillier \etal\, 1993) and might be used to place constraints on
physical parameters such as the volume filling factor (e.g., Oskinova
\etal\, 2001).

Therefore, we will consider a spherically symmetric and, on average, time independent
outflow that is a homogeneous mix of ``cool'' and ``hot'' gas in dynamical equilibrium.
For isothermal optically thin hot X-ray zones, the particular choice of $T_{\rm X}=10^7K$
tends to maximise the X-ray emission  in the {\it ROSAT} band (see Ignace \etal\, 2000).
We use the Raymond-Smith cooling function ($\Lambda_{\rm RS}$) for optically  thin plasmas
(Raymond \& Smith 1977). For WR~stars, a rough correction of $\Lambda_{\rm RS}$ for
non-solar abundances is given in Ignace \etal\, (2000). The ambient cool medium ($\le 10^5$ K)
is as described by the standard model and is optically thick to X-rays.

In the two-temperature approximation, the ratio of emission measures of hot and cool
components of the wind, ${\rm EM}_{\rm X}/{\rm EM}_{\rm w}$, is proportional to the ratio
of volumes filled by hot ($V_{\rm X}$) and cold ($V_{\rm w}$) material and to the
square of their density ratio $D_{\rm s}$. Thus, the ``filling factor'' is

\be
\fx=\,\frac{ {\rm EM}_{\rm X}}{{\rm {\rm EM}}_{\rm w}}\approx\,D_{\rm s}^2
\left (\frac{\mu_{\rm e}^{\rm w}}{\mu_{\rm e}^{\rm X}}\right )^2
\frac{V_{\rm X}}{V_{\rm w}},
\label{eq:few}
\ee

\noindent where  $\mu_{\rm e}^{\rm w}$ and $\mu_{\rm e}^{\rm X}$ are
the mean mass per free electron of the cool or standard wind
component.  The X-ray emission arises primarily from two-body
processes--collisional excitation and radiative recombination.
Therefore, the volume  emissivity \jnu\, (energy per unit time per unit
mass) is proportional to the square of the  density,

\be
4\pi\jnu\,=\,\fx \rho_{\rm w}^2\frac {\Lambda_{\nu}(\Tx)}
{\mu_{\rm e}^{\rm X}\mu_{\rm i}^{\rm X} m_{\rm H}^2}
\label{eq:jnu}
\ee

\noindent where $\mu_{\rm i}^{\rm X}$ is the mean molecular
weight per ion of the hot gas and $m_{\rm H}$ is the proton mass. 
$\Lambda_{\nu}$ (erg\,cm$^{3}$s$^{-1}$) is the cooling function
at the energies $E$ concerned and $\rho_{\rm w}$ is the density 
of the standard wind is

\be
\rho_{\rm w}(r)\,=\,\frac{\mdot}{4\pi r^2 v(r)}.
\label{eq:coneq}
\ee

\noindent By the definition of filling factor
in Eq.\,({\ref{eq:few}}), the total specific luminosity (erg\, s$^{-1}$)
emerging from the wind

\be
\Lx(E)\,=\,\fx \Lambda_\nu(T_{\rm X}) {\rm EM}_{\rm w}.
\label{eq:Lt}
\ee

\noindent This then is the monochromatic X-ray luminosity at
any instant in time, since our hypothesis is essentially that
the emission measure of X-ray emitting material (or correspondingly
the filling factor) is time variable.

It was pointed out by Cassinelli \etal\, (1996) that the X-rays from a
wind shock will be seen only after that shock has moved into a region
where the cool wind attenuation is optically thin to the X-rays.  As we
shall show, the wind is quite opaque at most X-ray energies up until
the flow has reached a substantial fraction ($\gtrsim 50\%$) of its
terminal speed.  Thus, we can assume that the X-rays emerge primarily
from distances where $v(r)\approx\vinf$.  For the wind attenuation,
K-shell absorption by metals in the cool wind is the dominant opacity source. 
Assuming photo-electric absorption by K-shell electrons

\be
\kappa_{\rm w}(E) = \sigma_\nu (E) / \mu_{\rm i}^{\rm w}\mH =
\frac{1}{\mu_{\rm i} m_{\rm H}}\,\sum_{\rm j} \,\frac{n_{\rm j}}{n_{\rm i}}\,\sigma_{\rm j}(E),
\label{eq:kapsig}
\ee

\noindent where $\mu_{\rm i}^{\rm w}$ is the mean molecular
weight per ion of the cool wind and $\sigma_{\rm j}(E)$ is the photo-absorption
cross-section. (Actually, the appropriate molecular weight to use is that per 
nucleus, but in hot star winds, there is no neutral gas, and so every nucleus 
is an ion.) The ratio $n_{\rm j}/n_{\rm i}$ defines the relative abundance by number 
for atomic species {\rm j}, where $n_{\rm i}$ is the number density of ions. The X-ray 
absorption (apart from prominent K-shell edges) can be represented roughly by 
a power-law in energy: $\kappaw(E)\approx \kappa_0 E^{-\gamma}$, with $\gamma$ 
in the range 2--3, depending on ionisation structure and chemical composition 
(see, e.g.  Hillier \etal\, 1993; Cohen \etal\, 1996). The parameter $\kappa_0$ 
is a constant that depends sensitively on the abundances.

To proceed, we now develop a phenomenological model in which the wind
inhomogeneities develop at random times near some inner boundary
$R_0$.  Then they propagate radially according to a monotonically
increasing velocity law $v( r )$. Let us introduce dimensionless
notations for velocity $u=v(r)/\vinf$, and for distance $x=r/R_0$.
Further, we normalise time to the characteristic time ${\cal T}_{\rm
f}=R_0/\vinf$, which is equal to the flow time in the specific case
$R_0=R_*$. Then for unitless time we have $\tm={\cal T}/{\cal T}_{\rm
f}$.

To avoid confusion we would like to clarify here, that further out,
 we shall use two descriptions for the distribution of the hot gas. In
the exospheric approximation described in Sect.\,3, we assume that the
hot material is present in the form of spherical shells.  The
exospheric approach with spherical shells are simplifying assumptions,
but the model does allow us to gain some insight into the nature of
X-ray variability from wind shocks.  In Sect.\,4, we allow for randomly
distributed wind shocks of arbitrary geometrical form (as long as the
hot zones are not exceedingly ``large'', to be defined later), and we
consider the full radiative transfer problem for the emergence of
X-rays from the wind flow.

For the motion of the hot zones (e.g., shells), we
consider an ensemble of zones filled by hot gas propagating in the
radial direction with constant velocity $u=1$.  The different zones in
our model are labelled by their times of appearance \tm\, at the
distance $x_0=1$.  The first zone is at distance $x_0=1$ at the moment
of time $\tm_0=0$. At time $\tm_k$ when shell number $k$ is crossing
$x_0=1$, the first zone is at distance $x^1(\tm_k)=\,x_0+ \tm_k$ with
$\tm_k=\sum_{i=1}^k \delta \tm_i$ and zone number $j$ is at distance
$x^j(\tm_k) =x_0+(\tm_k-\tm_j)$. Here, we choose to use the time
intervals $\delta \tm_i$ as drawn from uniformly random numbers in the
range $[0.5,\,1.5]$ multiplied by a dimensionless parameter $\langle T
\rangle$:

\be
\delta \tm_i\,=\,{\cal P}([0.5,1.5])\cdot\langle T \rangle,
\label{eq:vremya}
\ee

\noindent and so $\langle\tm_i\rangle = \langle T \rangle$.  
The parameter $\langle T \rangle$ is
in fact an average separation in time between the appearance of
subsequent zones and is measured in flow times ${\cal T}_{\rm f}$.
Further we will explore the influence of the average time separation
$\langle T \rangle$ on the character of the variability.

The parameter $\langle T \rangle$ is introduced because detailed models of X-ray
production demand the presence of an initial perturbation of the
stellar wind. For example, in models which depict X-ray production
mechanisms in O~stars (Feldmeier \etal\, 1997b), the existence of large
scale turbulence near the wind base is postulated.  As it will become
clear below, the average time separation $\langle T \rangle$ is the
only essential parameter of the models under consideration.  We thus
wish to emphasis that in principle, one can expect $\langle T \rangle$
to be an observable value, which could be inferred from the X-ray
emission light curve analysis as the average separation between the
peaks of intensity.

The emission measure of each zone is $\delta {\rm EM}=n_{\rm e}^{\rm
X}n_{\rm i}^{\rm X}\delta V$.  Using  Eq.\,(\ref{eq:coneq}), the
squared number density inversely scales with the fourth power of
distance:  $\nel\nion=\,n_0^2 \cdot x^{-4}$ , where 

\be n_0^2\,=\,D_{\rm s}^2\left( \frac{\mdot}{4\pi\vinf R_0^2}\right)^2 \frac{1}
{\mu_{\rm e}^{\rm x}\mu_{\rm i}^{\rm x} m_{\rm H}^2}.
\label{eq:ro0}
\ee

\noindent On the other hand, the volume of a zone $\delta V$ with fixed
radial thickness and solid angle grows with the square of distance $\delta
V\propto x^2$. This implies that  the contribution of a zone to the
total X-ray luminosity determined by its emission measure $\delta {\rm EM}$
has an overall decline that goes as $x^{-2}$.  This is a plausible
treatment when mass conservation for the hot shocked material is
assumed so that the continuity equation Eq.\,({\ref{eq:coneq}})
can be applied.

\section{Exospheric model for the time dependent X-ray emission of single stars}

\subsection{Modelling time dependent X-ray emission}

Let us assume that the observed X-ray emission arising from the hot gas
emerges only from radii exterior to the optical depth unity surface of
radius $r_1$ (exosphere), with X-rays at smaller radii completely attenuated. In
this approximation the X-ray luminosity has the proper scaling with
mass loss and other key wind parameters.  The exospheric radius $r_1$
of optical depth unity is  (Owocki \& Cohen 1999)

\begin{equation}
r_1(E)\,=\,\frac{\mdot}{4\pi\vinf}\kappa_{\rm w}(E).
\label{eq:r1}
\end{equation}

\noindent The emission measure of the cool material outside the
exosphere which is assumed to be transparent to the X-rays is given by

\begin{equation}
{\rm EM}_{\rm w} \approx 4\pi\,\int_{r_1}^\infty \,n_{\rm i}^{\rm w}n_{\rm e}
 (1-{\rm W}(r_1))\,r^2\,dr,
\label{eq:emwind}
\end{equation}

\noindent where ${\rm W}(r_1)$ is a dilution factor defined to be $\omega_{r_1}/4\pi$
where $\omega_{r_1}$ is the solid angle subtended by the exosphere  and
the parenthetical term accounts for geometric occultation by the
exospheric surface of radius $r_1$.

\begin{figure}[hbtp]
\centering
\mbox{\includegraphics[width=8.8cm]{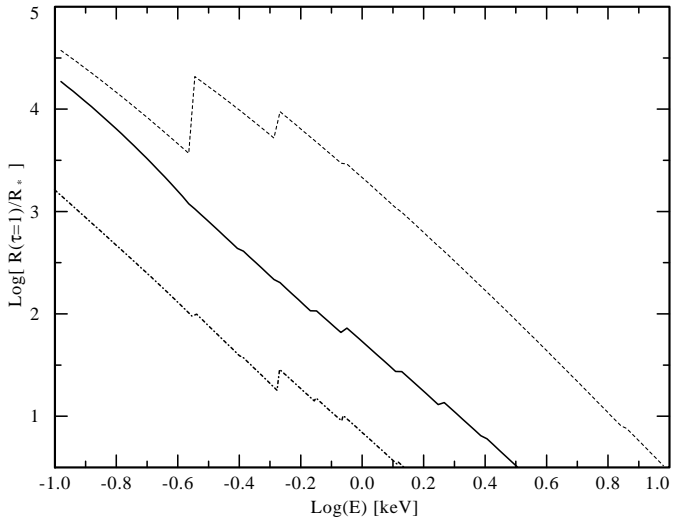}}
\caption [ ]{{Radius of optical depth unity as a function of X-ray energy.
Dashed-dotted line is for typical parameters of OI~stars, the solid line
is for typical WN~stars and the dashed line indicates WC~stars, which are
enhanced in C and O but deficient in N. Note the strong K shell edges due
to CNO elements}
\label{fig:f1} }\end{figure}

Let us make the simplifying assumption that X-ray emitting material is
present in the form of spherical shells.  All shells have the same
small fractional thickness ${\cal L}=\,\Delta R_{\rm shell}/R_0$ at a given
radius in the envelope. Recall that we assume isothermal X-ray
emitting zones and thus the parameter ${\cal L}$ is not to be
misconstrued as a cooling length but rather the geometrical fractional
thicknessof a shell filled with the hot gas.

In our  phenomenological exospheric model, X-ray emitting shells are seen only
beyond radius $r_1=x_1R_0$, so that effectively $x_0\equiv x_1$.  The
shells start their motion outwards in the atmosphere at quasi-random times
\tm$_k$. At that moment, the X-ray flux ($\delta F_{\rm X}$) is increased
by the new source. Then owing to the decrease in emission measure of the zone
with distance ($\delta {\rm EM}_{\rm X} \propto x^{-2}$ for
constant shell thickness), the incremental flux $\delta F_{\rm X}$
decreases. Thus, the total X-ray output of an ensemble of
X-ray emitting zones will be observed to vary with time.

The radius $x_1$ has a strong dependence on mass-loss rate, chemical
composition and energy (see Fig.~1).  This means that in the model under
consideration, the X-ray emitting zones will cross the surface of
optical depth unity at different distances for different energies.
Therefore their emission measures at the moment when they appear for an
external observer will be different.  Obviously then, one may expect
the character of variability to change as a function of wavelength
and for stars of different abundances.

The emission measure of an individual
 shell $j$ at time $\tm_k$ is

\be
\delta {\rm EM}_j\approx\,
4\pi R_0^3 n_{\rm e}^{\rm X}\,n_{\rm i}^{\rm X}\cdot{\cal L} \,\left[x_1(E)+
	(\tm_k-\tm_j)\right]^{-2}.
\label{eq:emx}
\ee

\noindent Combining Eqs.\,(\ref{eq:ro0}) and (\ref{eq:emx}), the
emission measure of the ensemble of  hot shells becomes
includegraphicsincludegraphics
\be
{\rm EM}_{\rm X}(E,\tm_{k_{\rm max}})\approx \,4\pi R_0^3 n_0^2{\cal L}
\cdot\sum_{k=1}^{k_{\rm max}}
\frac{1}{[x_1(E)+\tm_k]^2}.
\label{eq:emsum}
\ee

\noindent  Here the integral over the volume is replaced by summation
over all X-ray emitting shells outside the sphere of $x_1(E)$, and
$k_{\rm max}$ is the maximum number of shells under consideration in
the wind.  The value of $k_{\rm max}$ is chosen so that the most
distant shells make a negligibly small contribution to the emission
measure (i.e., $\delta EM_{\rm X}(x_{\rm max})\rightarrow 0$ for $x\gg
x_1$).  We choose $k_{\rm max}\sim 10^5$  which allows us to account
for the emission of a shell till it reaches a distance $x_{\rm
max}=k_{\rm max}\langle T \rangle\sim 10^5$.

The X-ray luminosity with time is $\Lx(E,t)={\rm
EM}_{\rm X}(E,\tm)\Lambda_\nu(\Tx)$.  Given that hot gas zones are a
function of radius only, and not of latitude or azimuth, a relatively
simple expression for the relative variability, $\sigma_{\rm L}/\Lx$,
can be derived.  Recognising that the X-ray emission from any given
zone drops rather steeply with radius, both the emission and
variability will be predominantly governed by zones that appear at the
$x_1$ radius.  Hence the fractional variability can be estimated based
on Poisson statistics.  The number density of zones above a given
radius $x$ scales as $1/x_1^2$; the volume in that vicinity scales as
$x_1^3$; so the number of zones near $x_1$ is $N \propto x_1$.
Consequently the relative variability scales as $\sigma_{\rm L}/\Lx
\propto 1/\sqrt{x_1}$.  Given that $x_1 \propto \kappa_\nu(E)$, we can
expect the relative variability to be a function of energy.  For
purposes of illustration, if $\kappa_\nu(E) \propto E^{-\gamma}$, then the
relative variability will be $\sigma_{\rm L}/\Lx \propto E^{\gamma/2}$,
which is an increasing function of energy.

\begin{figure}[hbtp]
\centering
\mbox{\includegraphics[width=8.8cm]{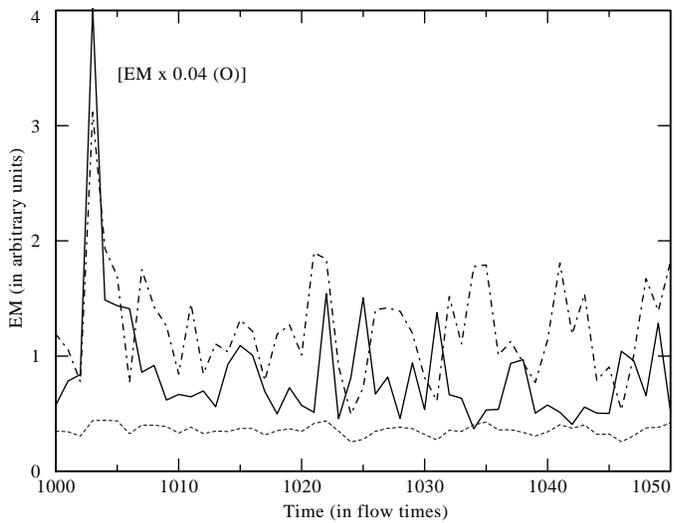}}
\caption [ ]{ Variations of emission measure with time for O, WN and
WC~stars.  The solid line represents the variation near 1.5 keV foincludegraphicsr a star
with an O~star chemical composition. The dash-dotted line is for a WN
star.  The dotted line is for a WC~star. Note that $\langle T
\rangle$=10 for these models. From the numerical simulations the ratios
of standard deviation $\sigma$ to the average value of emission measure
are $\sigma/\langle {\rm EM}\rangle=1.0$ for the O~star,
$\sigma/\langle {\rm EM}\rangle=0.38$ for the WN~star and
$\sigma/\langle {\rm EM}\rangle=0.13$ for the WC~star.  }
\label{fig:f2} 
\end{figure}

Figure\,2 shows changes in the emission measure of the shell ensemble,
computed using Eq.\,(\ref{eq:emsum}). We took the values of $x_1$
corresponding to 1.5\,keV for O, WN and WC stars. As is clear from the
figure, the character of variability depends on the chemical
composition. As can be seen from Fig.\,\ref{fig:f1}, the radius of
optical depth equal unity near 1\,keV for a WC star is located about an
order of magnitude further away in the wind than is the case for a WN
star.  This leads to an overall reduction of the X-ray emission, and
some suppression of the variability in WC~stars at this energy.  In
Sect.\,4 we revisit these simulations using the formal solution of
radiative transfer with angle dependent optical depth, instead of the
simplified exospheric approximation. But first we make some comments on
model constraints obtainable from filling factor data.

\subsection{Using filling factors to constrain model parameters}

In Ignace \etal\, (2000), lower limits to the filling factors of most
putatively single Galactic WR stars were determined from an analysis of
{\sc Rosat} All-Sky Survey (RASS) observations.  A natural question
arises: Is it possible to extract information about the spatial
distribution of hot, X-ray emitting gas by knowing \fx\, from RASS
passband observations in the framework of the model under
consideration?  Let us again assume that the fractional thickness
of the X-rayemitting layers ${\cal L}$ does not depend on distance.
By definition of Eq.\,(\ref{eq:few}) and using Eq.\,(\ref{eq:emwind})
and Eq.\,(\ref{eq:emsum}), we have:

\be
\fx\,=\,{\cal L}D_{\rm s}^2\left( \frac{\mu_{\rm e}^{\rm w}}{\mu_{\rm e}^{\rm X}}\right)^2
\frac{\sum_{k=1}^{N}\int_E (x_1+\tm_k)^{-2}dE}
{ \int_E x_1^{-2} dE}.
\label{eq:equ}
\ee

\noindent Here, we neglect the occultation term in
Eq.\,(\ref{eq:emwind}).  Note that $x_1=x_1(E)$ is implicit.  One
infers from Eq.\,(\ref{eq:equ}) that in fact there are two dependent
parameters in the problem: $\langle T \rangle$ and ${\cal L}$.  Let's
assume that it is possible to infer $\langle T \rangle$ from
observations of the variability of X-ray emission. Then making
assumptions about the density ratio $D_{\rm s}$, one may determine
${\cal L}$ if $L_{\rm X}$ (equivalently \fx) is known.  Therefore from
Eq.\,(\ref{eq:equ}), we may estimate ${\cal L}$ from known values of
{\fx}. In Table\,{\ref{tab:tab1}} we give typical values of ${\cal L}$
using $D_{\rm s}=4$ (the same value as in Baum \etal\, 1992 and Hillier
\etal\, 1993) for WN and WC stars.

So far, we have used the exospheric approximation to demonstrate that
the X-ray variability is strongly influenced by the opaque cool
material of the stellar wind and therefore (a) is a function of energy
and (b) depends on the chemical composition of the stellar wind (see
Fig.\,{\ref{fig:f2}}).  Also, it was shown that lower limits to filling
factors can be used to infer lower limits to the thickness of spherical
X-ray emitting shells (as illustrated in Tab.\,{\ref{tab:tab1}}), if
the average time interval between subsequent shocks $\langle T \rangle
$ is known from observations.

\begin{table}
\begin{center}
\caption{Typical values of ${\cal L}$ for {\rm WR}\,114 ({\rm WC5}) and {\rm WR}\,1 ({\rm WN5})$^a$
\label{tab:tab1}}
\begin{tabular}{ccc}\hline
\rule[0mm]{0mm}{4.0mm}
 & WR114: & WR1: \\
$\langle T \rangle$  &  $\fx=0.175 $ & $ \fx=0.0831 $ \\ \hline
\rule[0mm]{0mm}{4.0mm}
0.1	&  $6.2\cdot 10^{-3} $ & $ 5.1\cdot 10^{-3}$ \\
1.0	&  $4.5\cdot 10^{-2} $ & $ 2.5\cdot 10^{-2}$ \\
10.0    &  $4.4\cdot 10^{-1} $ & $ 2.0\cdot 10^{-1}$ \\ \hline
\end{tabular}
\end{center}
$^a$ Values of \fx\ and stellar parameters are from Ignace \etal\, (2000),
Hamann \& Koesterke (1998), and Koesterke \& Hamann (1995).
\end{table}


\section {Formal solution for X-ray emitting zones}

While it appears that the majority of WR and O~stars display basically
spherical winds on the large scale, there is compelling observational
evidence that the winds are clumped on small scales (e.g. L{\'e}pine \& Moffat 1999).
In L{\'e}pine \& Moffat (1999) a phenomenological  model was introduced,
wherein WR~winds are depicted as consisting of a large number of
randomly distributed, radially propagating, discrete wind emission
elements. This model was used to  simulate line profile variability
patterns in emission lines from a clumped wind. We adopt this approximation
for the X-ray producing region and consider the X-ray emission as
arising from analogous optically thin zones with temperature $T_{\rm X}$,
which as before are embedded in the cooler bulk wind flow. However, the zones
are no longer treated as spherical shells.

We treat these hot zones as  piecewise elements, presumably
produced by shocks. They expand adiabatically moving radially according
to a $\beta$-law for the velocity. Zones start their motion at randomly
distributed moments of time from random but uniformly distributed
positions at an imaginary spherical surface of radius $R_0$ -- the
minimal predicted radius for shocks to occur.

The main difference with the exospheric model from the previous section
is that the optical depth of cool material now depends not only on
radius but also on the spherical polar angle as seen by an observer.
Further we now compute a proper line-of-sight optical depth to each hot
emitting zone, in contrast to the exospheric approximation for which
hot zones are either completely attenuated, or completely
unattenuated.  To simplify the numerical simulation, we do require that
the emission measure of every zone $\delta {\rm EM}$ is the same at a
given radius and that the spatial size of each zone is small enough to
assume the wind attenuation to every point in the zone is approximately
constant (e.g., the zones cannot be hemispherical shells, since the
wind attenuation would strongly vary across such a structure).

In this section we adopt the usual cylindrical coordinate
system  (e.g. Lamers \etal\, 1987).  The coordinates are:  distance
from the star $x$, impact parameter for the line-of-sight $p$, and
distance along the line of sight $z$, all normalised to the stellar
radius. The angle between the line of sight and the radial vector is
$\theta=\arccos \mu$, with $\mu =1$ if the radial vector points to the
observer.  For any point in the wind, $x$, $p$ and $z$ are related by

\be
z\,=\,\mu x\,=\,\mu(p^2+z^2)^{1/2}
\ee

With our assumptions, the emission measure changes with distance from
the star as $\delta {\rm EM}_{\rm X}\,=\,\delta {\rm EM}_0(1/x)^2$,
with $\delta {\rm EM}_0$ the emission measure of a zone at distance
$x_0$.  The luminosity of each zone for which the optical depth of the
ambient wind is $\tau_{\rm w}$ is

\be
\delta L_{\nu}=\,\Lambda_{\nu}(\Tx)\,\delta {\rm EM}_{\rm X} \,e^{-\tau_{\rm w}}
\label{eq:dl}
\ee

\noindent and the luminosity of a hot gas zone occulted by the stellar core is
assumed to equal zero.  Applying the equation of continuity
Eq.\,(\ref{eq:coneq}) with Eq.\,(\ref{eq:kapsig}), we can derive the
wind optical depth to be

\be
\tau_{\rm w}=\frac{\sigma(E) \mdot}{4\pi R_0 \vinf \mu_{\rm i} m_{\sc H}} 
\int_z^{\infty} \frac{dz}{x^2 u(x)}\equiv\tau_0(E)\int _z^{\infty}
\frac{dz}{x^2 u(x)},
\label{eq:tauv}
\ee

\noindent where $x=\sqrt{p^2+z^2}$. Now the optical depth for each zone
is angle dependent.  For soft (0.4--2.4 keV) X-ray energies at radii
where the wind acceleration takes place, the optical depth is 
generally quite large (see Tab.\,\ref{tab:tab2}).

Although our numerical calculations can account for the X-ray emission
from zones at small radii, Table\,{\ref{tab:tab2}} indicates that their
contribution to the total X-ray flux will be quite negligible owing to
the rather extreme attenuation.  So, we simplify our calculations by
considering only those zones which have already reached the constant
velocity region in their motion throughout the wind, hence $u=1$. Using
Eq.\,(\ref{eq:tauv}), it is possible to obtain an analytic expression for
the optical depth (MacFarlane \etal\, 1991):

\be
\tau_{\rm w}=\, \tau_0(E)\int^{\infty}_{z}\frac{dz}{x^2}=
\,\frac{\tau_0(E)}{x}\frac{\theta}{\sin\theta}.
\label{eq:tautri}
\ee

\noindent  So, the optical depth depends
on the  location of a given zone in the wind as well as on the
energy.

\begin{table}
\begin{center}
\caption{Representative optical depths to an X-ray emitting zone propagating radially with
$\theta=45^{\circ}$ assuming a typical WN star$^a$ chemical composition.
\label{tab:tab2}}

\begin{tabular}{c|ccc}\hline
\rule[0mm]{0mm}{4.0mm}
u [v(r)/ \vinf] &  0.1         &   0.9           &  1.0        \\\hline
E [keV]      &                 &                 &             \\ \hline
\rule[0mm]{0mm}{4.0mm}
0.5	& $3.4\cdot 10^3$ & $3.5\cdot 10^2$ & $3.5\cdot 10^1$\\
1.0	& $5.5\cdot 10^2$ & $5.4\cdot 10^1$ & 5.5  \\
1.5          & $2.1\cdot 10^2$ & $2.0\cdot 10^1$ & 1.9  \\  \hline
\end{tabular}
\end{center}
$^a$ With $\vinf=2500\,\kmsec$ and $\mdot=10^{-4.4}$\msun yr$^{-1}$.
\end{table}

\begin{figure}[hbtp]
\centering
\mbox{\includegraphics[width=8.8cm]{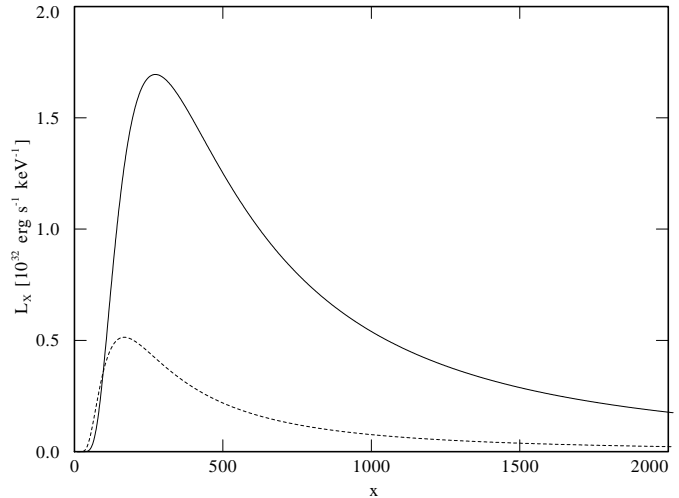}}
\caption [ ]{Change in monochromatic X-ray luminosity \Lx\ at 1\,keV
(solid line) and 1.2 keV (dashed line) of a single X-ray emitting zone with
distance $x$ in a WN~star wind. The zone moves along a
trajectory at $\theta=45^\circ$ in the envelope.
 }
\label{fig:f3}
\end{figure}

In Fig.~\ref{fig:f3} the change of X-ray flux for a single X-ray
emitting zone along with its motion through the flow is shown. The
luminosity grows rapidly with distance owing to the decrease in optical
depth by the overlying wind. In general the zones which are propagating
in directions close to $\cos\theta=1$ have higher luminosity due to the
fact that the optical depth is smaller in this direction.  The emission
rises to a peak followed by a decline.  At this point, the attenuation
is    negligible, and the decreasing emission is a consequence of the
inverse square fall-off of the emission measure with distance.  The
distance $x_{\sc peak}$ at which an X-ray emitting zone yields peak
emission depends strongly on the angle $\theta$. Further, noting that
$\delta{\rm EM}_{\rm X}\sim x^{-2}$ and $\tau_{\rm w}\sim x^{-1}$ it
can be shown (from an analysis of Eq.\,(\ref{eq:dl})) that the maximum
of the curve occurs when $\tau_{\rm w}(E,\theta)=2$.

Let us assume that a stellar wind contains numerous X-ray emitting
zones.  Clearly an ensemble of such zones will also produce time
variable X-ray emission. The rise in luminosity of a single zone is
much sharper than its decrease after peaking.  Suppose that at a given
time and a given energy, the X-ray luminosity of only one zone could
reach its maximum. Subsequently, the emission from this zone will be
decreasing, and until the next zone reaches its maximum, the emission
of the whole ensemble will drop.  Thus, in the regime of constant
expansion the separation between maxima of emission reflects the
average time separation $\langle T \rangle$. We may expect larger levels
of variability on a short time scale (order of flow times) for larger
values of ${\cal L}$. So, we may conclude that the observed lack of
variability suggests values of $\langle T\rangle\sim 1$ and therefore
values of ${\cal L}\sim 10^{-2}$, which are marginally consistent with
estimations of cooling lengths (e.g. Hillier \etal\, 1993). It is
necessary to point out here that the available observations so far have
not been capable of detecting such small fluctuations of X-ray flux.

\begin{figure}[hbtp]
\centering
\mbox{\includegraphics[width=8.8cm]{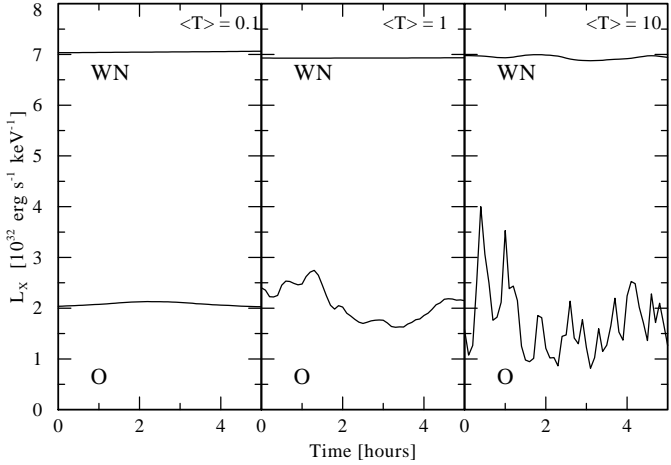}}
\caption [ ]{Change in narrow band X-ray luminosity \Lx\ at 1\,keV with time for
WR1 (WN5) (upper curve) and for  $\zeta$~Pup (O4If) (lower curve).
The average separation between X-ray emitting zones $\langle T \rangle$ grows
from the left panel to the right $\langle T \rangle=0.1,\,1,\,10$. The flow time
for the WN~type star is $t_{\rm f}\approx 0.15$~h and for the O~star is
$t_{\rm f}\approx 0.9$~h.}\label{fig:f4}
\end{figure}

To proceed further, let us recall that the total X-ray luminosity
depends on the random variable $\tm_i$. For the angular distribution of
hot zones, we select $\cos \theta$ as a uniform random variable in
interval $[-1, 1]$, and for the azimuth, $\phi_k$ is uniform random in
the interval $[0,2\pi]$.  It should be noted that although
$\cos\theta_k$ is uniform, $\tau_{{\rm w},k}$ is not.  The total
emission for an ensemble of zones is a summation over all contributors
with the appropriate attenuation:

\be
L_{\rm X}^{\rm tot}(E)=\,\Lambda_{\nu}(\Tx)\delta {\rm EM}_0
\sum_k^{k_{\rm max}}\frac{1}{(1+\tm_{\rm k})^2} e^{-\tau_{{\rm w},k}(E)}.
\label{eq:Ltottau}
\ee

\noindent The wind will be more opaque to X-rays at soft energies than
to those at hard energies. Therefore, we expect the phenomenological
picture wherein an X-ray emitting zone has peak flux at higher
energy earlier than it has peak flux at softer energy.

Using $\delta {\rm EM}_0$ as a parameter of the model, we avoid direct
references to the density of hot material or volume of the zones filled
by this material. In our model at each moment of time, only one X-ray
emitting zone is at a given distance from the inner boundary $R_0$.
This is similar to the propagation of subsequent spherical shells which
was described in the previous section in the framework of the
exospheric approximation. In Section\,3.2 we have shown that the
fractional thickness of a spherical layer ${\cal L}$ and $\langle T
\rangle$ are not independent parameters but coupled with each other
(see Eq.\,{\ref{eq:equ}).  To place constraints on $\delta {\rm EM}_0$,
let us assume now that $\delta {\rm EM}_0$ is the same as the emission
measure of a spherical shell with thickness ${\cal L}$ located at
distance $R_0$ with corresponding density  $n_0$ as in
Eq.\,(\ref{eq:ro0}).

The number of hot gas zones and their initial emission measure $\delta
{\rm EM}_0$ is set by the average separation in time $\langle T\rangle$.  
The assumption of the constant filling factor leads to the
correlation between average time interval and initial emission measures
of hot zones.  This means that small hot zones with small separation in
time should produce small changes in X-ray flux. On the other hand,
large sized hot zones produce changes in X-ray flux with a more
significant amplitude.  To complete the picture, it is necessary to
bear in mind that the degree of variability also depends on chemical
composition.

Fig.~\ref{fig:f4} presents numerical simulations of short-scale time
variability of the monochromatic X-ray luminosity at 1\,keV for two stars with
different cool wind opacities, namely O~stars and WN~stars.
Clearly, the character of variability strongly depends on the average
time separation $\langle T \rangle$ and the spectral type of the star.
The mass loss rates of WN~stars are much higher then those of O stars;
therefore, we attribute the difference in attenuation between WR~stars
and O~stars as the main cause for different character of variability
shown in Fig.~\ref{fig:f4} owing to the fact that $x_1$ is much bigger
for WN stars than for O~stars.

The changes in luminosity seen in Fig.\,{\ref{fig:f4}} are, in fact,
due to the superposition of light curves for many zones propagating in
different directions similar to those shown in Fig.\,{\ref{fig:f3}}.
The average separation between subpeaks on the curves from
Fig.\,{\ref{fig:f4}} reflects the average time between two subsequent
zones having peaks in their luminosity. However, when $\langle T
\rangle$ is small, the flux fluctuations are negligible (order $\sim$
few per\,cent) and simply cannot be resolved.  The interesting question
to address is what is the plausible range for the parameter $\langle T
\rangle$.  Observed fluctuations in X-ray emission for $\zeta$ Ori
(O4f) (increase in the count rate of $\approx$ 30\%  for 2 days) and
$\zeta$ Pup (O9.5Ia) (modulations with a period of 16.7 hours and
amplitude $\leq$ 10\%, Bergh{\" o}fer \& Schmitt 1994, Bergh{\" o}fer
\etal\, 1996) suggest rather small values of $\langle T \rangle$
($\leq  1$ flow time), assuming that the X-rays form in wind shocks.

As shown by our numerical simulations, on the time scale of several
hours, the variability of X-ray luminosity might be negligible for
stars of spectral type WN.  The reasons for this apparent lack of
variability are that in addition to small size and short time
separation of hot zones, the total optical depth for the indicated
energies is quite large in the WN~star winds. As can be seen from
Eq.\,({\ref{eq:Ltottau}}), the exponential term $e^{-\tau_{\rm w,i}}$
suppresses the diference in values of \Lx\,. Although not shown in
Fig.\,\ref{fig:f4}, our modelling reveals that the X-ray luminosity of
WC~stars is almost constant due to their quite opaque stellar winds. As
obvious from the figure, even the transparent winds of O~stars may not
demonstrate detectable levels of variability in X-rays on short time
scales of about 1 hour in the case of small $\langle T \rangle$. 

\begin{figure}[hbtp]
\centering
\mbox{\includegraphics[width=8.8cm]{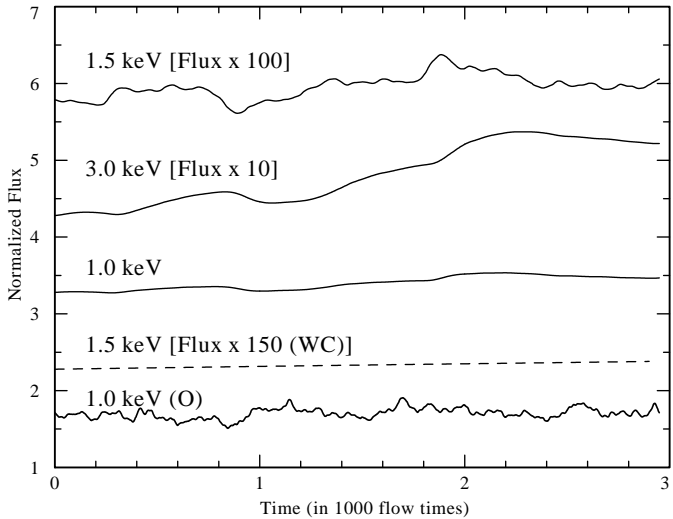}}
\caption [ ]{Change in monochromatic X-ray flux at 1\,keV, 1.5\,keV and
3\,keV with time for WR1 (WN5) (solid lines), for  $\zeta$~Pup (O5Ia)
(lowest curve) and a typical WC~star (dashed line). The average
separation between X-ray emitting zones is $\langle T \rangle\,=\,1$.
}
\label{fig:f6}
\end{figure}

Although on a time-scale of hours, the 
X-ray variability may be quite small, the dynamical flow time $r/v(r)$ for
distances $r\geq 100 R_*$ is some $10^5$ seconds, of order a day.  So
we have performed simulations of changes in flux over longer time
intervals. The results of this calculation are shown in
Fig.\,{\ref{fig:f6}}.  Surprisingly, even for $\langle T \rangle=1$,
the variability becomes significant ($\sim$ 10\%) on time scales of
thousands of hours. In this case, the separation between peaks in the
light curves do not reflect the time separation between X-ray emitting
zones. To understand the character of this variability let us consider
Eq.\,(\ref{eq:Ltottau}).  The time dependent luminosity is governed by
two uniformly distributed random variables $\delta \tm_i$,  and $\cos
\theta_i$. Let us assume, for simplicity, that zones are launched
within constant time intervals equal to unity and concentrate on random
variations in $\tau_{\rm w}$.  Then $\delta \tm_i=1$ and from
Eq.\,({\ref{eq:Ltottau}})

\be
\Lx^{\rm tot}\propto\,\sum_k^\infty\frac{1}{(1+k)^2}\ e^{-\tau_{w,k}}.
\label{eq:L_1}
\ee

\noindent
From Eq.\,({\ref{eq:tautri}}), $\tau_{w,k}\propto \tau_0\theta_k/k\sin
\theta_k$ and $\cos\theta_k$ is a random variable from $[-1, 1]$. Thus,
when the number of realisations $k$ is large enough, even for rather
opaque winds with $\tau_0\gg 1$, the variance of the flux  is
significant.

The likelihood of detecting variability of X-ray flux increases
drastically at hard energies. This is due to the strong dependence of
the optical depth on energy. To illustrate this, if $\kappa_{\rm
w}(E)\sim E^{-\gamma}$ then using Eq.~(\ref{eq:L_1}) and neglecting the
second order terms, the standard deviation of the X-ray luminosity becomes
$\sigma_{\rm L}\sim E^{\frac{\gamma}{2}}(\mdot/\vinf)^{-\frac{1}{2}}$.
\noindent This is an increasing function of energy and decreasing
function of the stellar wind density.  As numerical simulations show,
in the case of random times \tm$_{\rm i}$, the general trend of the
dependence of variance on energy and density is an increasing function
of energy. The only condition is that the variance of emission measures
of the hot zones should be smaller than the variance of the optical
depths associated with them.

Fig.\,\ref{fig:f5} represents the ratio of the standard deviation
$\sigma_{\rm L}$ to the average value of the X-ray luminosity as a
function of energy, computed using Eq.\,(\ref{eq:Ltottau}).  Apart from
ionisation edges, Fig.\,\ref{fig:f5} clearly shows that the
relative variability is an increasing function of energy.
Fig.\,\ref{fig:f5} confirms the basic scaling of variability with
parameter $x_1$ that was obtained under the 
exospheric approximation with $\sigma_{\rm L}/\Lx\propto 1/\sqrt{x_1}$.
Presumably, this will fail for large values of $\langle T \rangle \sim
x_1$, because then there are relatively few hot zones in the
outflow. At very low $\langle T \rangle$ one hits another regime
where the flow approaches homogeneity.

As seen in Fig.\,\ref{fig:f5} for $\langle T \rangle=1$, the amplitude
of variability for O~stars can reach as much as 80\% near 1.5 keV but
only 10\% at softer energies around 0.5 keV.  At the same time, the
relative variability is only a few percent near 1 keV for the
WN~stars.  For WC~stars, the level of X-ray variability is negligible.
Therefore, it is not surprising that with such low sensitivity  X-ray
emission detectors such as {\sc Rosat} PSPC (0.2--2.4 keV) or \sc
Einstein} IPC (0.2--4.0 keV), the detected X-ray flux of WR~stars
appeared to be fairly constant X-ray sources.  

\begin{figure}[hbtp]
\centering
\mbox{\includegraphics[width=8.8cm]{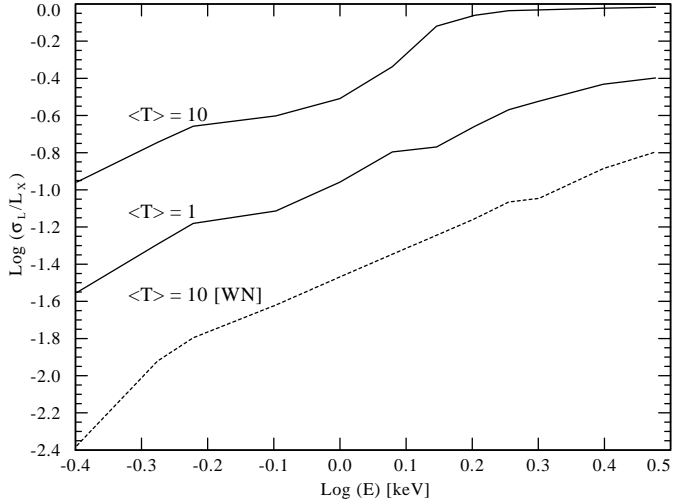}}
\caption [ ]{Dependence of the ratio of standard deviation $\sigma_{\rm
L}$ to the value of the X-ray luminosity$\langle L_{\rm X} \rangle$
versus energy for typical abundances of O~stars (solid lines) and for
the average chemical composition of a typical WN star (dashed line).  }
\label{fig:f5}
\end{figure}

\section{Summary and conclusions}

We present simulations of the expected X-ray variability for early-type
stars in the framework of a shock model for X-ray production.  We
assumed that the optically thin hot X-ray emitting material is embedded
in a cool X-ray absorbing stellar wind, described by the standard
model. For such a medium we used the concept of filling factor, which
is the ratio of emission measures of X-ray emitting and absorbing
material.  Assuming an isothermal X-ray emitting hot gas, the filling
factor is proportional to the ratio of volumes filled by hot and cool
components and to the square  of the ratio of densities of hot and cool
material.  To examine the basic processes of X-ray emission and
absorption we considered two different models.

To begin, we employed the exospheric approximation with angle
independent optical depth.  In this approximation we used a model of
the stellar wind where hot gas is present in the form of spherical
shells propagating with constant velocity. The emission of such an
outflow is time dependent. This allowed us to derive a expression 
restricting the fractional thickness of a spherical shell
emitting X-rays by using filling factors.  The lower limits on filling
factors for isothermal X-ray emitting material were derived in Ignace
\etal\, (2000). Using these data, we found lower limits to the
thickness of the spherical shells filled by hot gas to produce the
observed level of X-ray emission. It was shown that the thickness of
an isothermal spherical shell can be inferred from the analysis of the
X-ray light curve and the observed filling factor.

Further, the model of an envelope consisting of a number of individual
radially propagating X-ray emitting zones was developed by analogy with
stochastic wind models. The total X-ray luminosity of such an
atmosphere was obtained by using the formal solution of the radiation
transfer equation for angle dependent optical depth. We did not place
any constraints on the form, volume or density of the zones of hot gas
except to demand that the total emission measure of each zone is the
same as the emission measure of a spherical shell at the same distance
from the stellar core, and that the zone must be relatively ``small''.
This allows us to restrict the number of independent parameters when
performing the numerical simulations. The rationale for doing this was
that broadband X-ray luminosities  are known for most hot stars.
Therefore, any combination of parameters describing the distribution of
hot gas should provide the observed level of X-ray emission for each
particular star as its first priority.

The two major simplifying assumptions of our modelling are those of
isothermal X-ray sources and smooth cool material. We may speculate
that including radiatively cooling sources will lead to a decrease in
the level of variability. On the other hand, taking into account the
clumpy structure of the background wind will lead to an increase of the
variability of X-ray output. A detailed consideration of these
effects are a matter of future work.

Despite its appoximative nature, the analysis described here revealed
the basic properties of variability of X-ray emission for early type
stars. Some of the main points are:

\begin{enumerate}

\item The apparent lack of short time-scale variability (order of an
hour) of X-ray emission cannot be considered as a deficiency of shock
models for X-ray production.

\item We may expect stochastic variability on long time-scales
(thousands of flow times), especially for optically thick winds.

\item The level of X-ray variability depends on chemical composition
and density of the stellar wind and differs for stars of different
spectral classes even for similar mechanisms of X-ray production.  It
is governed by the opacity of the cool material and is substantially
lower for the more opaque winds of WR~stars and particularly WC~stars.

\item The dependence of the wind opacity with energy means that
the X-ray emission may be highly variable in the part of the spectrum,
where the cool material is optically thin and practically constant in
optically thick parts. Therefore the hard energies are specially apt
for detection of variability of X-ray emission.  Recall that existing
claims of X-ray variability show differences between soft and hard
passbands (e.g., possibly detected for $\zeta$~Ori by Bergh{\"o}fer \&
Schmitt 1994).

\item Whether  X-ray variability is detected or not for a given energy
provides valuable information about the spatial distribution and
properties of X-ray emitting material, and if variability is detected,
its dependence on energy would be especially telling of the wind
structure.

\end{enumerate}

\begin{acknowledgements}

This research was supported by a PPARC grant (LMO, JCB) and NASA grants
NAG5-9964 (RI) and NAG5-9226 (JPC). We acknowledge discussions with
K.~Gayley, M.~Hendry and S.~Rauzy, and we thank the referee
L.~Koesterke for comments beneficial to this paper.

\end{acknowledgements}

\end{document}